\documentclass[onecolumn,prd,amsmath,amssymb,showkeys]{revtex4}

\usepackage{amsmath}
\usepackage{graphicx}
\usepackage{graphics}
\usepackage{dcolumn}
\usepackage{bm}
\usepackage{hyperref}

\def\BEq{\begin{equation}}
\def\EEq{\end{equation}}
\def\BEqA{\begin{eqnarray}}
\def\EEqA{\end{eqnarray}}
\def\BEn{\begin{enumerate}}
\def\EEn{\end{enumerate}}
\def\BWT{\begin{widetext}}
\def\EWT{\end{widetext}}

\def\dag{\dagger}
\def\adj{{^{\dag}}}

\newcommand{\cev}[1]{\reflectbox{\ensuremath{\vec{\reflectbox{\ensuremath{#1}}}}}}

\begin{document}

\title{Canonical quantization of the dark positive-energy Dirac field and time asymmetry}

\author{Andrei Galiautdinov}
\email{ag1@uga.edu}
 \affiliation{
Department of Physics and Astronomy, 
University of Georgia, Athens, GA 30602, USA}

\date{\today}
\begin{abstract}

We perform canonical quantization of the single-component, 
spin-zero field that was introduced by Dirac in 1971 and recently 
suggested as a candidate for dark matter by Bogomolny. 
The massive and massless cases are treated separately. 
Since in the massive case only positive-frequency modes 
are normalizable and regarded as physical, the mode expansion 
for the field involves annihilation operators only, making the 
quantization procedure particularly simple. The corresponding 
Hamiltonian turns out to be unambiguous, with 
no need for normal ordering. The positive-energy requirement
imposed on the second-quantized system leads to equally 
acceptable Bose and Fermi choices for particle statistics.
This suggests a simple extension of original Dirac's 
theory in which Bose and Fermi single-component positive-energy 
Dirac fields are combined into a doublet whose members can transform 
into each other. The model includes a Landau-Anderson-Higgs type
potential that allows spontaneous selection of the direction in the 
internal bose-fermi space.
The bosonic sector of the theory hints at the possibility of 
a dark, background spacetime condensate that could endow 
the universe with its cosmological temporal asymmetry.    
In the massless case of Dirac's original theory, we explore the possibility 
of allowing the field expansion to involve both positive- and 
negative-frequency modes. This leads to the anticommutation relations 
for creation and annihilation operators associated with the negative 
energy solutions, resulting in supersymmetric behavior of the 
single-component field in the ultrarelativistic limit. Finally,
we speculate on the possibility for the positive-energy Dirac
particles to obey some exotic (such as non-abelian, Clifford)
statistics in which the particles are neither created nor destroyed.
\end{abstract}


\maketitle


\section{Introduction}

The concept of dark matter, which was invoked by Zwicky 
\cite{zwicky1933,zwicky1937} to account for the anomalous 
motion of galaxies near the edge of the Coma Cluster, had 
been the subject of great theoretical interest for almost a 
hundred years. Since Zwicky's original proposal, the list of 
astrophysical phenomena that require dark matter for their 
explanation has steadily been expanding and now includes 
at least half a dozen entrees, such as formation and evolution 
of galaxies, galaxy rotation curves, mass position in galactic 
collisions, gravitational lensing, anisotropies of the cosmic 
macrowave background, 
as well as the evolution and structure of the Universe as a whole
\cite{cirelli2024}.
The current estimate consistent with the standard Lambda-CDM 
cosmological model puts the dark matter contribution 
to the mass-energy content of the Universe at about 26.8\%, 
which is disproportionally large in comparison to the meager 
4.9\% contrubuted by ordinary matter \cite{jarosik2011,ade2013}. 
Because its presence 
may only be detected gravitationally, traditional models attribute 
dark matter to a yet-to-be-discovered class of weakly interacting 
particles, but the exact nature and composition of such particles 
remains unknown. It is therefore not surprising that the ``dark 
matter problem'' is currently regarded as one of the most 
important unsolved problems in physics.

In this regard, the work by Dirac from 1971 on the so-called 
positive-energy relativistic wave equation 
\cite{dirac1971,dirac1972,dirac1978,dirac1982} 
(which is colloquially known as the ``new'' Dirac equation) 
acquires a special significance.
Unlike the more familiar equation of 1928 which describes 
spin-1/2 
particles \cite{dirac1928}, Dirac's new equation 
describes a single-component spin-0 field, $\psi(x; q_1,q_2)$, 
which
depends not only on the spacetime coordinates, $x^{\mu}$, 
but also on two additional parameters, $q_{1}$, $q_{2}$, 
which represent two auxiliary quantum mechanical degrees 
of freedom
(see Eq.\ (\ref{eq:NDE}) below). The new Dirac 
equation has recently attracted renewed attention 
\cite{bogomolny2024,cirilo-lombardo2023,cirilo-lombardo2024a,
cirilo-lombardo2024,cirelli2024,galiautdinov2024},
as it exhibits some remarkable properties, chief among them 
being the positivity of the energy 
of its single-particle modes and the lack of a mathematically 
consistent procedure for the introduction of coupling to the 
electromagnetic field. This latter property (which was originally 
viewed as a drawback) is especially profound as it
prevents observation of the new Dirac field by standard 
astronomical methods 
and points towards its potential importance in gravitational 
physics as a viable 
candidate for dark matter \cite{bogomolny2024}. 
In the context of new physics, such as in the search for dark 
matter, supersymmetry, and quantum gravity, Dirac's new 
formulation may play an important role in opening up new 
avenues for exploring theoretical formulations that extend 
physics beyond the Standard Model \cite{bogomolny2024}. 
By ensuring a well-defined 
and physically meaningful solution space, Dirac's treatment 
of positive-energy states could offer a pathway for theories 
that reconcile quantum mechanics with gravity, potentially 
leading to new insights into the fundamental nature of 
spacetime and quantum fields.

The mathematical theory 
underlying Dirac's new equation was reviewed in sufficient detail in 
Refs.\ \cite{bogomolny2024} and \cite{mukunda1982}, to which the 
reader is directed. Here, in Section \ref{sec:background}, we 
provide a brief summary, and then 
turn to canonical quantization of 
$\psi(x; q_1,q_2)$, which to our knowledge had not been 
previously performed [though the adopted procedure is analogous 
to the one used by Sudarshan and Mukunda \cite{sudarshan1970} 
in the case of the infinite-component Majorana field \cite{majorana1932}.]
Due to the absence of the negative-frequency modes, the field 
quantization avoids the use of creation operators, which 
automatically eliminates the need for normal ordering in the 
resulting Hamiltonian. While curious in itself, and notwithstanding 
its potential relevance to the cosmological constant problem, 
the physical implication of this 
result is not immediately obvious, since any realistic theory 
involving Dirac's new field would necessarily involve gravity, 
which is the only entity to which $\psi(x; q_1,q_2)$ can 
couple directly. In what follows,
we restrict consideration to flat spacetime only, leaving the 
question of quantization in the presence of gravity to future 
study.

\section{Theoretical Background}

\label{sec:background}

The ``new'' positive-energy relativistic wave equation 
\cite{dirac1971,dirac1972,dirac1978,dirac1982} has the form,
\BEq
\label{eq:NDE}
({\gamma}^{\mu} \partial_{\mu} - m) q\psi(x;q_1,q_2) = 0,
\EEq 
subject to the consistency condition,
\BEq
\label{eq:consistencyCondition}
(\partial^{\mu} \partial_{\mu} + m^2)\psi(x;q_1,q_2) = 0,
\EEq
where $q$ is the column, ${q}=(q_1, q_2, q_3, q_4)^{\rm T}$, 
with $q_3=-i\partial/\partial q_1$, $q_4=-i\partial/\partial q_2$, and
the gamma matrices,
\BEq
\gamma^0
\equiv
\begin{pmatrix}
0 & 0 & 1 & 0 \cr
0 & 0 & 0 & 1 \cr
-1 & 0 & 0 & 0 \cr
0 & -1 & 0 & 0
\end{pmatrix},
\quad
\gamma^1
\equiv
\begin{pmatrix}
-1 & 0 & 0 & 0 \cr
0 & 1 & 0 & 0 \cr
0 & 0 & 1 & 0 \cr
0 & 0 & 0 & -1
\end{pmatrix},
\quad
\gamma^2
\equiv
\begin{pmatrix}
0 & 1 & 0 & 0 \cr
1 & 0 & 0 & 0 \cr
0 & 0 & 0 & -1 \cr
0 & 0 & -1 & 0
\end{pmatrix},
\quad
\gamma^3
\equiv
\begin{pmatrix}
0 & 0 & -1 & 0 \cr
0 & 0 & 0 & -1 \cr
-1 & 0 & 0 & 0 \cr
0 & -1 & 0 & 0
\end{pmatrix},
\EEq
satisfy the anticommutation relations,
\BEq
{\gamma}^{\mu} {\gamma}^{\nu} 
+ {\gamma}^{\nu} {\gamma}^{\mu} 
= 
-2\eta^{{\mu}{\nu}},
\quad
\eta^{{\mu}{\nu}} = {\rm diag}(+,-,-,-),
\quad
{\mu},{\nu}=0,1,2,3.
\EEq
Considering propagating plane wave solutions in the form,
\BEq
\psi_{p}(x;q_1,q_2)
=  
u(p;q_1,q_2) e^{-ip_{\mu}x^{\mu}},
\quad
p_{\mu} = (p_0,-p_x,-p_y,-p_z),
\EEq
Dirac's new equation comprises a system of four equations \cite{dirac1971},
\begin{align}
\label{eq:system1}
[(p_0+p_z)q_3+(p_x-im)q_1-p_yq_2]u &=0,
\\
\label{eq:system2}
[(p_0+p_z)q_4-(p_x+im)q_2-p_yq_1]u &=0,
\\
\label{eq:system3}
[(p_0-p_z)q_1+(p_x+im)q_3-p_yq_4]u &=0,
\\
\label{eq:system4}
[(p_0-p_z)q_2-(p_x-im)q_4-p_yq_3]u &=0,
\end{align}
three of which are independent, whose solutions are the modes,
\BEq
\label{eq:modes}
u(p; q_1,q_2)
=
\exp\left\{
-\frac{(m+ip_x)q_1^2 - 2ip_y q_1q_2+(m-ip_x)q_2^2}{2(p_0+p_z)}
\right\},
\EEq
with
\BEq
\label{eq:1particleEnergies}
p_{0} \equiv \pm\omega_{\bm{p}},
\quad
\omega_{\bm{p}}\equiv + \sqrt{m^2 + \bm{p}^2}>0,
\EEq
of which only the positive-energy ones are normalizable for $m>0$.
This can be seen by inspecting the simplest wave function corresponding 
to zero momentum,
\BEq
\label{eq:DiracSimplestSolution}
u = \exp\left[-(m/p_0) (q_1^2 +q_2^2)\right],
\EEq
which for $p_0 =-m$ is indeed non-normalizable, since
$\int_{-\infty}^{\infty} dq_1 dq_2 |u|^2 \rightarrow \infty$.
As a result, the general physically acceptable solution of the new Dirac 
equation for massive case, $m> 0$, can be written as the superposition 
of positive-frequency modes only,
\begin{align}
\label{eq:expansion}
\psi(x;q_1,q_2)
&=
\int \frac{d^3{\bm p}}{\sqrt{(2\pi)^3 2 p_0}} \,
a(p) \, u(p; q_1,q_2)  \, e^{-i(p_0 t-{\bm p} {\bm x})},
\quad
aq=qa,
\end{align}
subject to $p^0>0$ and normalization \cite{mukunda1982}, 
\begin{align}
\label{eq:modeNormalization}
(\psi_{p},\psi_{p'}) 
&\equiv
\int \frac{d^3{\bm x}}{(2\pi)^3} \int dq_1 dq_2 \;
\psi_{p}^{*}(x; q_1,q_2)
\,
\left(q_1^2+q_2^2+q_3^2+q_4^2\right)
\,
\psi_{p'}(x; q_1,q_2)
\nonumber \\
&=
\int \frac{d^3{\bm x}}{(2\pi)^3}  \int dq_1 dq_2 \;
\psi_{p}^{*}\,\tilde{q}q\,\psi_{p'}
\nonumber \\
&=
-\int \frac{d^3{\bm x}}{(2\pi)^3}  \int dq_1 dq_2 \;
\psi_{p}^{*}\,\tilde{q}\gamma^0 \gamma^0 q\,\psi_{p'}
\nonumber \\
&=-\int \frac{d^3{\bm x}}{(2\pi)^3}  \int dq_1 dq_2 \;
\psi_{p}^{*}\,\bar{q} \gamma^0 q\,\psi_{p'}
\nonumber \\
&=
2p^0 \delta(\bm{p}-\bm{p}'),
\end{align}
where $a(p)$ are the expansion coefficients to be interpreted as the 
annihilation operators.

We note in passing, that in the low-energy (non-relativistic) limit, 
the system of equations (\ref{eq:system1}) through (\ref{eq:system4}) 
has the form,
\begin{align}
\label{eq:system1nr}
\left[\left(1+\frac{p_z}{m}\right)q_3+\left(\frac{p_x}{m}-i\right)q_1
-\frac{p_y}{m}q_2\right]u &=0,
\\
\label{eq:system2nr}
\left[\left(1+\frac{p_z}{m}\right)q_4-\left(\frac{p_x}{m}+i\right)q_2
-\frac{p_y}{m}q_1\right]u &=0,
\\
\label{eq:system3nr}
\left[\left(1-\frac{p_z}{m}\right)q_1+\left(\frac{p_x}{m}+i\right)q_3
-\frac{p_y}{m}q_4\right]u &=0,
\\
\label{eq:system4nr}
\left[\left(1-\frac{p_z}{m}\right)q_2-\left(\frac{p_x}{m}-i\right)q_4
-\frac{p_y}{m}q_3\right]u &=0,
\end{align}
with the solution,
\BEq
\label{eq:modesnr}
\psi_{p}(t, \bm{x};q_1,q_2)
=
e^{-imt}e^{-i\left(\frac{{\bm p}^2}{2m}t-{\bm p}{\bm x}\right)}
e^{
-\frac{1}{2}\left(1-\frac{p_z}{m}\right)\left(q_1^2+q_2^2\right)
}
e^{-\frac{i}{2}\left\{\frac{p_x}{m}
\left(q_1^2-q_2^2\right)-2\frac{p_y}{m} q_1q_2 \right\}},
\EEq
which obeys the modified Schr\"{o}dinger equation
of standard quantum mechanics. On the other hand, in the high-energy 
(ultrarelativistic) limit corresponding to $m\rightarrow 0$, we have,
\begin{align}
\label{eq:system1ur}
[(p_0+p_z)q_3+p_xq_1-p_yq_2]u &=0,
\\
\label{eq:system2ur}
[(p_0+p_z)q_4-p_xq_2-p_yq_1]u &=0,
\\
\label{eq:system3ur}
[(p_0-p_z)q_1+p_xq_3-p_yq_4]u &=0,
\\
\label{eq:system4ur}
[(p_0-p_z)q_2-p_xq_4-p_yq_3]u &=0,
\end{align}
with the solution ({\it cf.} \cite{ousmanemanga2013a}),
\BEq
\psi_{p}(x;q_1,q_2)
=
e^{-ip_{\mu}x^{\mu}}
e^{-\frac{i}{2}\frac{p_x(q_1^2-q_2^2)-2p_yq_1q_2}{p_0+p_z}},
\EEq
without any real-valued exponential prefactor. This makes ultrarelativistic 
(massless) wave functions normalizable, at least in principle (though some 
subtleties involved), which will be explored in Section 
\ref{sec:includingNegativeEnergyWaveFunctions}. 


\section{Quantization Procedure}
\label{sec:quantization}

In flat spacetime the proposed action for the dark Dirac field is 
(\cite{galiautdinov2024,ahner1975,ahner1976,
castillo2011,ousmanemanga2013}; {\it cf.}\ \cite{sudarshan1970}),
\begin{align}
\label{eq:actionANDlagrangian}
S_{\rm D}=\int d^4 x \int dq_1dq_2 \; {\cal L}_{\rm D},
\quad
{\cal L}_{\rm D}
=
-i 
\left\{
\frac{1}{2}
\left[
\bar{\Psi} {\gamma}^\mu (\partial_\mu \Psi)
-
(\partial_\mu \bar{\Psi})  {\gamma}^\mu \Psi
\right]
-m \bar{\Psi} \Psi
\right\},
\end{align}
with
\BEq
\Psi = q\psi,
\quad
\Psi\adj = \psi\adj\tilde{q},
\quad
\bar{\Psi} = \Psi\adj \gamma^0 = \psi\adj\tilde{q} \gamma^0
=\psi\adj \bar{q},
\quad
\bar{q}=\tilde{q}\gamma^0,
\quad
\tilde{q}=(q_1, q_2, q_3, q_4),
\EEq
where tilde denotes matrix transposition. Variation of the action with 
respect to the ``big'' $\bar{\Psi}$ and $\Psi$, which following 
\cite{castillo2011,ousmanemanga2013} we regard as the 
independent dynamical variables of the theory, recovers the new Dirac 
equation (\ref{eq:NDE}), together with its adjoint, 
\BEq
\bar{\Psi}(\cev{\partial}_\mu {\gamma}^\mu + m) = 0.
\EEq
The two canonically conjugate momenta are then,
\begin{align}
{\cal P}
&= 
\frac{\delta \cal{L}_{\rm D}}{\delta (\partial_0 \Psi)}
=
-\frac{i}{2}\bar{\Psi}\gamma^0
=
\frac{i}{2}{\psi}\adj \tilde{q}
=\frac{i}{2} ({\psi}\adj q_1, {\psi}\adj q_2, {\psi}\adj q_3, {\psi}\adj q_4),
\\
\bar{\cal P} 
&= 
\frac{\delta \cal{L}_{\rm D}}{\delta (\partial_0 \bar{\Psi})}
=
+\frac{i}{2}\gamma^0 {\Psi}
=
\frac{i}{2}\gamma^0 q{\psi}
=
\frac{i}{2} 
\begin{pmatrix}
+q_3 \psi \cr +q_4 \psi \cr -q_1 \psi \cr -q_2 \psi
\end{pmatrix},
\end{align}
leading to the Hamiltonian density,
\begin{align}
\cal{H}_{\rm D}
&=
{\cal P} \partial_0{\Psi} + (\partial_0{\bar{\Psi}}) \bar{\cal P} - {\cal L}_{\rm D}
=
-\frac{i}{2}
\left[
\bar{\Psi}\gamma^0 \partial_0 \Psi
-
(\partial_0 \bar{\Psi})\gamma^0 \Psi
\right]
+
i 
\left\{
\frac{1}{2}
\left[
\bar{\Psi} {\gamma}^\mu (\partial_\mu \Psi)
-
(\partial_\mu \bar{\Psi})  {\gamma}^\mu \Psi
\right]
-m \bar{\Psi} \Psi
\right\},
\end{align}
which on the mass shell assumes two equivalent forms,
\begin{align}
\label{eq:H1}
\cal{H}_{\rm D}
&=
-\frac{i}{2}
\left[
\bar{\Psi}\gamma^0 \partial_0 \Psi
-
(\partial_0 \bar{\Psi})\gamma^0 \Psi
\right] \quad \text{(on-shell)},
\end{align} 
and
\begin{align}
\cal{H}_{\rm D}
&=
\frac{i}{2}
\left[
\bar{\Psi}(\gamma^k \partial_k - m) \Psi
-
\bar{\Psi}(\cev{\partial}_0 \gamma^0 + m )\Psi
\right] \quad \text{(on-shell)}.
\end{align} 
For the purpose of quantization we regard the field as being in the 
Heisenberg picture and use the first option, Eq.\ (\ref{eq:H1}), to 
simplify the calculation.
We then have, for $m>0$,
\begin{align}
\label{eq:expansion}
\psi(x)
&=
\int 
\frac{d^3{\bm p}}
{\sqrt{(2\pi)^3 2 p_0}} \, a(p) \, u(p) \, 
e^{-i(p_0 t-{\bm p} {\bm x})},
\\
{\psi}{\adj}(x)
&=
\int 
\frac{d^3{\bm p}}{\sqrt{(2\pi)^3 2 p_0}} \, 
a\adj(p) \, {u}^{*}(p) \, e^{i(p_0 t-{\bm p} {\bm x})},
\\
\partial_0 \psi(x)
&=
\int 
\frac{d^3{\bm p}}
{\sqrt{(2\pi)^3 2 p_0}}\, (-ip_0) \, a(p) \, u(p) \, 
e^{-i(p_0 t-{\bm p} {\bm x})},
\\
\partial_0 {\psi}\adj(x)
&=
\int 
\frac{d^3{\bm p}}
{\sqrt{(2\pi)^3 2 p_0}} \, (+ip_0) \, a\adj(p) \, {u}^{*}(p)  \, 
e^{i(p_0 t-{\bm p} {\bm x})}  ,
\end{align}
whose substitution into Eq.\ (\ref{eq:H1}) gives,
\begin{align}
\label{eq:Hamiltonian}
{H}_{\rm D}
&=
-\frac{i}{2}
\int d^3 {\bm x} \int dq_1dq_2 \;
\left[
\bar{\Psi}\gamma^0 \partial_0 \Psi
-
(\partial_0 \bar{\Psi})\gamma^0 \Psi
\right]
\nonumber \\
&=
\frac{1}{2}
\int \frac{d^3 {\bm p} d^3 {\bm p}'}{\sqrt{2p_0 2p'_0}}
\,
\left(-
\int \frac{d^3 {\bm x}}{(2\pi)^3}\int dq_1dq_2 \;
e^{i(p_0 t-{\bm p} {\bm x})}
\, {u}^{*}(p)  \, \bar{q}\gamma^0 \, q \, u(p')\, 
e^{-i(p'_0 t-{\bm p}' {\bm x})}
\right)
(p'_0+p_0) \,  a\adj(p) \, a(p') 
\nonumber \\
&=
\frac{1}{2}
\int \frac{d^3 {\bm p} d^3 {\bm p}'}{\sqrt{2p_0 2p'_0}}
\,
\left(-
\int \frac{d^3 {\bm x}}{(2\pi)^3}\int dq_1dq_2 \;
\psi_{p}^{*}\,\bar{q} \gamma^0 q\,\psi_{p'}
\right)
(p'_0+p_0) \,  a\adj(p) \, a(p') 
\nonumber \\
&=
\frac{1}{2}
\int d^3 {\bm p} d^3 {\bm p}'
\, \delta(\bm{p}-\bm{p}') \,
\frac{2p^0 (p'_0+p_0)}{\sqrt{2p_0 2p'_0}} \,  a\adj(p) \, a(p') 
\nonumber \\
&=
\int d^3 {\bm p} \, p_0 \,  a\adj(p) \, a(p) \,,
\end{align} 
with {\it no need for normal ordering}. 

To arrive at an acceptable particle interpretation we require the 
Hamiltonian ${H}_{\rm D}$ to be positive-definite, which 
automatically allows two equally acceptable possibilities 
({\it cf}.\ \cite{sudarshan1970}): either commutation relations,
\BEq
\label{eq:commRelations}
[a(p),a\adj(p')]=\delta^3(\bm{p}-\bm{p}'),
\quad
[a(p),a(p')]=0,
\quad
[a\adj(p),a\adj(p')]=0,
\EEq
or anticommutation relations,
\BEq
\label{eq:anticommRelations}
\{a(p),a\adj(p')\}=\delta^3(\bm{p}-\bm{p}'),
\quad
\{a(p),a(p')\}=0,
\quad
\{a\adj(p),a\adj(p')\}=0,
\EEq
indicating possible existence of both bosonic and fermionic positive-energy 
Dirac particles. [Remark: To prevent potential confusion, we 
point out that our bosonic creation and annihilation operators 
bear no relation to the ones given in Eq.\ (8.1) of Dirac's 
original paper \cite{dirac1971}. Dirac's ``creation'' and 
``annihilation'' operators were a useful auxiliary 
construct of his single-particle theory. They operated on the internal 
space of states and were formally used to switch from ``position'' 
representation of internal wave functions expressed in terms of 
$q_1$ and $q_2$, to the Fock representation in terms of another 
set of variables, $\eta_1=(q_1-iq_3)/\sqrt{2}$ and 
$\eta_2=(q_2-iq_4)/\sqrt{2}$, that were associated with two abstract 
harmonic oscillators. See \cite{bogomolny2024} for elaboration of this 
point.]

\section{Including negative-energy modes}
\label{sec:includingNegativeEnergyWaveFunctions}

Here we explore the possibility of incorporating into 
the theory the negative-frequency modes. 
We do that by restricting the domain of system's 
internal variables to a finite region, say, $q_1,q_2 \in [-\ell,\ell]$, taking 
the limit $m\rightarrow 0$, and imposing (if needed) periodic boundary conditions 
(or changing the topology of the $q$-space in some other fashion), 
making the negative-frequency modes normalizable. 
This procedure is partially supported by considering the simplest 
case: Dirac's simplest solution (\ref{eq:DiracSimplestSolution}) is formally 
normalizable for $p_0<0$ when the domain is reduced and periodicity 
conditions are imposed, even for $m>0$. There is some possibility that we 
may lose self-adjointness of various ``internal'' quantum-mechanical operators, 
and that periodicity conditions may not be consistently imposed, but those 
complication, we posit, can likely be handled by various self-adjoint 
extension techniques \cite{gitman2012}. For now we proceed formally, 
ignoring potential mathematical subtleties.

Then, the field has the expansion
({\it cf.} Eq.\ (\ref{eq:expansion}); 
here $v=v(p;q_1,q_2)$ denotes negative-frequency modes),
\begin{align}
\psi(x)
&=
\int \frac{d^3{\bm p}}{\sqrt{(2\pi)^3 2 \omega_{\bm p}}}
\left[
 a(p) \, u(p) \, e^{-i(\omega_{\bm p} t-{\bm p} {\bm x})}
+
b\adj(p)v(p) \, e^{i(\omega_{\bm p} t-{\bm p} {\bm x})}
\right],
\end{align}
which, after some algebra, results in
\begin{align}
\label{eq:HamiltonianFull}
{H}_{\rm D}
&= 
\int d^3 {\bm p} \, \omega_{\bm p} \,  a\adj(p) \, a(p)
\nonumber \\
&-
\int 
\frac{d^3 {\bm p} d^3 {{\bm p}'}}
{\sqrt{2\omega_{\bm p} 2\omega_{{\bm p}'}}}
\,
(\omega_{\bm p}+\omega_{{\bm p}'})
\left(-\frac{1}{2}
\int \frac{d^3 {\bm x}}{(2\pi)^3}
e^{-i(\omega_{\bm p} t-{\bm p} {\bm x})}
\, \bar{V}(p)  \, \gamma^0 \, V(p')\, 
e^{i(\omega_{{\bm p}'} t-{\bm p}' {\bm x})}
\right)
 \, b(p) \, b\adj(p') 
\nonumber \\
&-
\int 
\frac{d^3 {\bm p} d^3 {\bm p}'}
{\sqrt{2\omega_{\bm p} 2\omega_{{\bm p}'}}}
\,
(\omega_{\bm p}-\omega_{{\bm p}'})
\left(-\frac{1}{2}
\int \frac{d^3 {\bm x}}{(2\pi)^3}
e^{-i(\omega_{\bm p} t-{\bm p} {\bm x})}
\, \bar{V}(p)  \, \gamma^0 \, U(p')\, 
e^{-i(\omega_{{\bm p}'}-{\bm p}' {\bm x})}
\right)
 \, b(p) \, a(p') 
\nonumber \\
&+
\int 
\frac{d^3 {\bm p} d^3 {\bm p}'}
{\sqrt{2\omega_{\bm p} 2\omega_{{\bm p}'}}}
\,
(\omega_{\bm p}-\omega_{{\bm p}'})
\left(-\frac{1}{2}
\int \frac{d^3 {\bm x}}{(2\pi)^3}
e^{i(\omega_{\bm p} t-{\bm p} {\bm x})}
\, \bar{U}(p)  \, \gamma^0 \, V(p')\, 
e^{i(\omega_{{\bm p}'} t-{\bm p}' {\bm x})}
\right)
 \, a\adj(p) \, b\adj(p'),
\end{align} 
where we introduced the notation, $U = qu, V = qv$.
If we assume that the negative-energy modes are ``regularized''  
in the above mentioned sense, then we get,
\begin{align}
\label{eq:HamiltonianFullFormal}
{H}_{\rm D}
&= 
\int d^3 {\bm p} \, \omega_{\bm p} \,  
\left[
a\adj(p) \, a(p) - b(p) \, b\adj(p)
\right].
 \end{align}
This leads to several possibilities.

The first one is to impose anticommutation relations on the $b$-modes, 
\BEq
\{b(p),b\adj(p')\}=\delta^3(\bm{p}-\bm{p}'),
\quad
\{b(p),b(p')\}=0,
\quad
\{b\adj(p),b\adj(p')\}=0,
\EEq
making them fermionic, while using bosonic commutation relations 
(\ref{eq:commRelations}) for the $a$-modes, together with the 
commutation relations between the two,
\BEq
[a(p),b(p')]=0,
\quad
[a(p),b\adj(p')]=0,
\EEq
That would lead to a positive-definite Hamiltonian with a positively-divergent 
term,
\begin{align}
{H}^{(1)}_{\rm D}
&= 
\int d^3 {\bm p} \, \omega_{\bm p} \,  
\left[
a\adj(p) \, a(p) + b\adj(p) \, b(p) 
\right]
+
\int d^3 {\bm p} \, \omega_{\bm p}\delta({\bm 0}).
 \end{align}
One may speculate about a strange possibility here, where depending on 
the energy regime (slow vs ultrarelativistic) the $b$-modes, initially 
non-normalizable and physically ``frozen'', get ``activated'' as the 
energy tends to infinity. [This property of the new Dirac field 
in the ultrarelativistic regime may have connection 
to the proposal expressed in \cite{cirilo-lombardo2023} 
(also see \cite{sanchez2019}) with regard to the inclusion of 
fermionic coordinates in the quantum description of the light-cone.]
Our system then would exhibit supersymmetric behavior,
with supersymmetric operators defined by (e.\ g., \cite{kaku1993}),
\BEq
Q(p)\equiv b\adj(p) a(p) + a\adj(p)b(p), 
\EEq
which commute with the Hamiltonian,
\BEq
[Q(p),H^{(1)}_{\rm D}(p)]=0,
\EEq
and satisfy
\BEq
\{Q(p),Q\adj(p) \} = \frac{2}{\omega_{{\bm{p}}}} H^{(1)}_{\rm D}(p).
\EEq

Other options include either working exclusively within the fermionic sector 
of the theory in all regimes, or working within the fermionic sector in the 
ultrarelativistic regime only (in which case the strange scenario would be 
associated with the 
initially bosonic $a$-modes switching their statistics upon the ``activation'' 
of the fermionic $b$-modes).

Finally, there is an option of simply redefining the $b$-operators,
$b \leftrightarrow b\adj$, and working exclusively within the bosonic sector, 
while leaving the Hamiltinian as is.
In this case we would get,
\begin{align}
{H}^{(2)}_{\rm D}
&= 
\int d^3 {\bm p} \, \omega_{\bm p} \,  
\left[
a\adj(p) \, a(p) - b\adj(p) \, b(p)
\right],
 \end{align}
containing the negative-energy contribution, 
which may prove useful for dealing with the cosmological constant problem 
\cite{zeldovich1968,weinberg1989}. 

\section{Extending original Dirac's theory}

Returning to the $m>0$ case, the dual statistical nature  
expressed by relations (\ref{eq:commRelations}) and 
(\ref{eq:anticommRelations}) raises the
question as to which of the two possibilities for statistics obeyed by
the positive-energy Dirac particles Nature actually 
chooses and how. We hypothesize that the choice is made spontaneously
by an analogue of the Anderson-Higgs mechanism familiar from the 
Standard Model. This calls for an extension of Dirac's original 
single-component equation (somewhat in the spirit of, but different from, 
e.\ g., 
Refs.\ \cite{horvathy2008},\cite{okninski2014},\cite{cirilo-lombardo2024a}), 
by considering a doublet consisting of two dark spin-0 fields with opposite 
statistics,
\BEq
\Psi 
=
\begin{pmatrix}
\Psi_1\cr
\Psi_2
\end{pmatrix}
\equiv
\begin{pmatrix}
q\psi_1\cr
q\psi_2
\end{pmatrix},
\quad
\bar{\Psi} 
=
\begin{pmatrix}
\bar{\Psi}_1, \bar{\Psi}_2
\end{pmatrix}
\equiv
\begin{pmatrix}
{\psi}^\dag_1 \bar{q}, {\psi}^\dag_2\bar{q}
\end{pmatrix},
\EEq
where $\psi_1$ describes spin-0 bosons and $\psi_2$ describes spin-0 fermions. 
Different directions in this internal doublet space, characterized by the 
corresponding values of the mixing angle, would then correspond 
to universes with different properties, 
possibly leading to cosmological evolutions with varying degrees of temporal 
asymmetry. In our extension, we choose a Lagrangian that allows for 
self-interaction and transmutation between the doublet members, 
for example, in the form,
\begin{align}
\label{eq:extendedLagrangian}
{\cal L}^{\text{(ext)}}_{\rm D}
=
-i 
\left\{
\frac{1}{2}
\left[
\bar{\Psi} {\gamma}^\mu (\partial_\mu \Psi)
-
(\partial_\mu \bar{\Psi})  {\gamma}^\mu \Psi
\right]
- 
\bar{\Psi} 
\begin{pmatrix}
m_1 & g \cr 
g & m_2 
\end{pmatrix}
\Psi
- \frac{i \lambda}{2} (\bar{\Psi} \Psi)^2
\right\},
\end{align}
where $g$ and $\lambda$ are the real-valued coupling constants, with $g$
being (speculatively) responsible for the corresponding internal space oscillation. Variation of the action 
involving (\ref{eq:extendedLagrangian}) results in the system of equations,
\begin{align}
({\gamma}^{\mu} \partial_{\mu} - m_1) \Psi_1 -g\Psi_2 -i\lambda(\bar{\Psi}\Psi)\Psi_1 = 0,
\quad
({\gamma}^{\mu} \partial_{\mu} - m_2) \Psi_2 -g\Psi_1 -i\lambda(\bar{\Psi}\Psi)\Psi_2 = 0,
\end{align}
or, using $\bar{q}q=2i$, in terms of ``little'' $\psi$s,
\begin{align}
({\gamma}^{\mu} \partial_{\mu} - m_1) q\psi_1 -gq\psi_2 
+\lambda({\psi}^\dag_1{\psi}_1 + {\psi}^\dag_2{\psi}_2)q\psi_1 = 0,
\quad
({\gamma}^{\mu} \partial_{\mu} - m_2) q\psi_2 -gq\psi_1 
+\lambda({\psi}^\dag_1{\psi}_1 + {\psi}^\dag_2{\psi}_2)q\psi_2 = 0,
\end{align}
generalizing Eq.\ (\ref{eq:NDE}). A model with the ``wrong'' sign for either of the 
masses (or both), say,
$m_1=-M_1<0$, $m_2=+M_2>0$, or $m_1=-M_1<0$, $m_2=-M_2<0$,
would then allow for some specific direction in the doublet space (together 
with the corresponding universe) to be spontaneously chosen.


\section{Discussion}
\label{sec:discussion}

The interpretation in terms of creation and annihilation operators adopted 
in Section \ref{sec:quantization} that ensures positivity of system's energy 
may not be the only possibility. Some time ago, Nayak and Wilczek 
\cite{nayak1996,wilczek1998} put 
forward a proposal for a non-abelian cliffordonic statistics based on 
projective representations of the permutation group 
\cite{karpilovsky1985,hoffman1992}, in 
which the corresponding particles were neither created nor destroyed, 
but, instead, permuted. Originally, Nayak and Wilczek's statistics was intended 
for a description of quasiparticles in the quantum Hall effect, but was 
later used in some approaches to spacetime quantization (see, e.\ g.,
\cite{baugh2001,finkelstein2001,galiautdinov2002a,galiautdinov2002b,baugh2003};
for recent work in that direction \cite{marks2022}).
The main idea behind the latter attempts was to provide a unified description
for quantum fields and space-time (the so-called ``quantum-field-spacetime unity'') 
in which the universe was regarded as a collection of fundamental building blocks 
(let's call them chronons, for lack of a better word) whose permutations underlie 
the fundamental physical processes going on in nature. 
[For thermodynamic implications of Clifford statistics see \cite{galiautdinov2002c};
for the appearence of a spin-orbit coupling not present in the Standard Model see
\cite{galiautdinov2002a,galiautdinov2002b}.] In the present context, one may 
also try to investigate the possibility of nonabelian statistics for particles obeying 
Dirac's new equation (provided careful attention is paid to the critical position 
expressed in Ref.\ \cite{read2003} regarding the subject). 
This speculation hinges on the curious form, Eq.\ (\ref{eq:modes}), of the mode 
functions of the new Dirac theory, which allows topological modification of 
the $q$-manifold and clearly distinguishes the $z$-direction in the 
momentum space (especially in the deep, non-relativistic regime, 
Eq. (\ref{eq:modesnr}), where the $p_z$ term becomes irrelevant, 
making the system effectively two-dimensional). It is also supported by the large
body of work on anyonic statistics in relation to equations of Dirac-Majorana type  
(see, e.\ g., \cite{duval2001,horvathy2002,horvathy2004}, and related to our 
subject Ref.\ \cite{levy-leblond1967}). 
Needless to say, the quantum-mechanical operators $a(p)$ and $a\adj(p)$ 
employed in such an exploration may end up carrying a very different physical 
interpretation.

Returning to the commutation and anticommutation relations 
(\ref{eq:commRelations}), (\ref{eq:anticommRelations}), 
let us make a brief remark that whenever we encounter bosons we  
should also check for the possibility of a phase transition. 
In the present context that dictum naturally directs our attention to 
cosmology. In the recent work \cite{cirilo-lombardo2023, cirilo-lombardo2024} 
it was demonstrated that entanglement and coherent states applied to 
quantum description of deSitter space and black holes can lead to asymmetric 
time evolution and that the equations of Dirac-Majorana type may play a role 
in that description. Since (as far as we know) gravity is the only entity with 
which the new Dirac field is allowed to interact, the corresponding phase 
transition may turn out to be of either BEC or BCS type (the latter, 
superconductor-like, case could be due to Cooper pairing of the fermionic 
spin-0 particles via the second order virtual exchange). The relevant question 
then is: 
Could it be possible that the time asymmetry present in the macroscopic world 
\cite{davies1977,penrose1979,hawking1985,gell-mann1994,
zeh2011, ellis2013,vilenkin2013,cortes2018} is the manifestation of the two 
systems (gravity plus the new dark Dirac field) working together to form a 
spacetime condenstate that enforces a preferred set of initial conditions and 
makes different parts of the universe ``propagate'' in the same temporal direction?

Recall that in general relativity a spacetime manifold is not just 
a manifold equipped with a Lorentzian metric. Rather, it is a 
Lorentzian manifold that has an additional structure, called time 
orientation, that has to be postulated separately (essentially, 
introduced by hand). The resulting time-oriented manifold then
acquires a globally consistent notion of the direction of time -- the 
causal structure that, at each spacetime event, distinguishes 
between the past and the future. Somewhat more formally, 
relativistic spacetime is a four-tuple, $(M, g, \nabla, T)$, in 
which $ (M, g, \nabla)$ is a Lorentzian manifold with metric 
$g$ and torsion-free connection 
$\nabla$, and $ T $ is a smooth timelike vector field, satisfying 
$g(T, T) > 0$ (it is this field that gives spacetime its 
time-orientation). General relativity does not say anything 
about the physical nature of $T$. Instead, the field $T$ is 
typically ``provided'' by other, non-gravitational parts of physics, 
such as, e.\ g., thermodynamics or quantum mechanics,
and it is expected that, ultimately, it will emerge from a more 
fundamental theory. What we are suggesting here, 
is that, in the absence of the fundamental theory, and by 
analogy with how in general relativity energy and matter 
influence spacetime curvature, the time-orientatability of the 
spacetime manifold may also be affected by its material 
content. It may very well happen that the energy-momentum 
vector of the bose-condensed positive-energy Dirac particles 
plays the role of such time-orienting vector field.

\appendix

\section{Modifying variational procedure to include the Klein-Gordon constraint}
\label{appendixA}

The Klein-Gordon consistency constraint, Eq.\ (\ref{eq:consistencyCondition}), may be
incorporated into our variational procedure by adding the extra term involving
Lagrange multiplier, $\lambda$,
\BEq
{\cal L}_{\rm constraint} 
= 
\lambda \left(\partial^{\mu}\partial_{\mu}\psi+m^2\psi\right)
=
\partial^{\mu}(\lambda \partial_{\mu}\psi)
-
(\partial^{\mu}\lambda)(\partial_{\mu}\psi)+m^2\lambda\psi,
\EEq
to the Lagrangian in (\ref{eq:actionANDlagrangian}), and requiring that
the little field $\psi$ and the big field $\Psi\equiv q\psi$ be varied independently.
Variation of $\lambda$ then immediately recovers the constraint 
(\ref{eq:consistencyCondition}), while variation of the little $\psi$ shows that
$\lambda$ obeys its own Klein-Gordon equation,
\BEq
\left(\partial^{\mu}\partial_{\mu}+m^2\right)\lambda = 0.
\EEq

\section{Mandelstam reformulation of the new Dirac equation}
\label{appendixB}

Here, for the sake of completeness, we show how the impossibility 
of minimal coupling to the electormagnetic field reappears in Mandelstam's 
\cite{mandelstam1962} formulation of the new Dirac theory.

In the presence of electromagnetic potentials, $A_\mu$, Dirac's new equation 
for the gauge-invariant path-dependent field (with $\Pi$ being the defining path
running from spatial infinity),
\BEq
\phi(x;q_1,q_2;\Pi)
=
\psi(x;q_1,q_2)e^{-ie\int_{-\infty}^{x}d\xi^{\mu}A_{\mu}(\xi)},
\EEq
takes the form,
\BEq
\label{eq:MNDE}
({\gamma}^{\mu} \partial_{\mu} - m) q\phi(x;q_1,q_2) = 0,
\EEq 
where the partial derivatives $\partial_{\mu}$ acting on $\phi(x;q_1,q_2)$ no
longer commute. Direct calculation then gives,
\BEq
\label{eq:dmdnphi}
\partial_{\mu}\partial_{\nu} \phi 
=
-ie(\partial_{\mu}A_{\nu})\phi + \text{(terms symmetric w.r.t. the 
$\nu\leftrightarrow \mu$ interchange)}.
\EEq
Applying the operator $({\gamma}^{\nu} \partial_{\nu} + m)$ 
to Eq.\ (\ref{eq:MNDE}) and using Eq.\ (\ref{eq:dmdnphi}) leads to the equation,
\begin{align}
\left[-\left(\eta^{\mu\nu}\partial_{\mu}\partial_{\nu}+m^2\right)
-ie\sigma^{\mu\nu}\partial_{\mu}A_{\nu}\right]q\phi=0,
\end{align}
where $\sigma^{\mu\nu}=(1/2)[\gamma^{\mu},\gamma^{\nu}]$.
The consistency argument similar to the one presented in Sec\ 5 of 
Ref.\ \cite{bogomolny2024} then leads to the requirement 
$\partial_{\mu}A_{\nu}=0$, for $\mu\neq \nu$, 
or, $F_{\mu\nu}=0$ (Faraday's tensor), thus preventing
electromagnetic coupling, in agreement with Dirac's 
original theory.


\begin{thebibliography}{}

\bibitem{zwicky1933}
F.\ Zwicky, The red shift of extragalactic nebulae,
Helvetica Physica Acta 6, 110 (1933).

\bibitem{zwicky1937}
F.\ Zwicky,
On the Masses of Nebulae and of Clusters of Nebulae, 
Astrophys.\ J.\ 86, 217 (1937).

\bibitem{cirelli2024}
M.\ Cirelli, A.\ Strumia, J.\ Zupan,
Dark Matter, 
arXiv:2406.01705 [hep-ph]

\bibitem{jarosik2011}
N.\ Jarosik, et al.,
Seven-year Wilson microwave anisotropy probe (WMAP) observations: Sky maps, systematic errors, and basic results,
Astrophys.\ J.\ Suppl.\ 192 (2), 14 (2011). 

\bibitem{ade2013}
P.\ A.\ R.\ Ade, N.\ Aghanim, C.\ Armitage-Caplan, et al.\ (Planck Collaboration) , Planck 2013 results. I. Overview of products and scientific results, 
Astr.\ Astrophys.\ 1303, 5062 (2013).

\bibitem{dirac1971}
P.\ A.\ M.\ Dirac, 
A positive-energy relativistic wave equation,
Proc.\ R.\ Soc.\ Lond.\ A 322,
435 (1971).


\bibitem{dirac1972}
P.\ A.\ M.\ Dirac, 
A positive-energy relativistic wave equation. II,
Proc.\ R.\ Soc.\ Lond.\ A 328, 1 (1972).



\bibitem{dirac1978}
P.\ A.\ M.\ Dirac, 
{\it Directions in Physics} (John Wiley \& Sons, New York, 1978).


\bibitem{dirac1982}
P.\ A.\ M.\ Dirac, 
Pretty mathematics,
Int.\ J.\ Theor.\ Phys., 21, 603 (1982) .

\bibitem{dirac1928}
P.\ A.\ M.\ Dirac, 
The quantum theory of the electron,
Proc.\ R.\ Soc.\ Lond.\ A 117, 610 (1928).

\bibitem{bogomolny2024}
E.\ Bogomolny,
Positive-Energy Dirac Particles and Dark Matter,
Universe, 10, 222 (2024); arXiv:2406.01654 [hep-th].


\bibitem{cirilo-lombardo2023}
D.\ J.\ Cirilo-Lombardo and N.\ G.\ Sanchez, 
Coherent states of quantum spacetimes for black holes and de Sitter spacetime,
Phys.\ Rev.\ D 108, 126001 (2023).

\bibitem{cirilo-lombardo2024a}
D.\ J.\ Cirilo-Lombardo and N.\ G.\ Sanchez, 
Quantum-Spacetime Symmetries: A Principle of Minimum
Group Representation,
Universe, 10, 22 (2024).

\bibitem{cirilo-lombardo2024}
D.\ J.\ Cirilo-Lombardo and N.\ G.\ Sanchez, 
Entanglement and Generalized Berry Geometrical Phases in Quantum Gravity,
Symmetry, 16(8), 1026 (2024).

\bibitem{galiautdinov2024}
A.\ Galiautdinov,
Positive-energy Dirac particles in cosmology,
arXiv:2406.08699 [gr-qc]



\bibitem{mukunda1982}
N.\ Mukunda, H.\ van Dam, L.\ C.\ Biedenharn,
{\it Relativistic Models of Extended Hadrons Obeying a 
Mass-Spin Trajectory Constraint}, 
Lectures in Mathematical Physics at the University of Texas 
at Austin, edited by A.\ B\"{o}hm and J.\ D.\ Dollard
(Springer-Verlag, Berlin, Heidelberg, 1982).

\bibitem{sudarshan1970}
E.\ C.\ G.\ Sudarshan and N.\ Mukunda,
Quantum Theory of the Infinite-Component Majorana Field and the
Relation of Spin and Statistics,
Phys.\ Rev.\ D 1, 571 (1970).

\bibitem{majorana1932}
E.\ Majorana, 
Teoria relativistica di particelle con momento intriciseco arbitrario,
Nuovo Cimento, 9, 335 (1932).

\bibitem{ousmanemanga2013a}
A.\ Ousmane Manga, A.\ Moussa, A.\ Aboubacar, N.\ V.\ Samsonenko,
Two-Component Form of the New Dirac Equation,
Adv.\ Studies Theor.\ Phys., 7(7), 319 (2013).

\bibitem{ahner1975}
H.\ Ahner, 
Gravitation and positive-energy wave equation,
Phys.\ Rev.\ D 11, 3384 (1975).


\bibitem{ahner1976}
H.\ Ahner, 
Action principle for spin-1/2 wave equation,
Phys.\ Rev.\ D 13, 250 (1976).


\bibitem{castillo2011}
J.\ E.\ Castillo H., A.\ H.\ Salas,
A Covariant Relativistic Formalism for the New Dirac Equation,
Adv.\ Studies Theor.\ Phys., 5(8), 399 (2011).

\bibitem{ousmanemanga2013}
A.\ Ousmane Manga, N.\ V.\ Samsonenko, A.\ Moussa,
Lagrangian Formalism for the New Dirac Equation,
Adv.\ Studies Theor.\ Phys., 7(3), 141 (2013).


\bibitem{gitman2012}
D.\ M.\ Gitman, I.\ V.\ Tutin, B.\ L.\ Voronov,
{\it Self-adjoint Extensions in Quantum Mechanics}
(Springer, New York, 2012).

\bibitem{sanchez2019}
N.\ G.\ Sanchez, 
New Quantum Structure of Space-Time,
Gravit.\ Cosmol., 25, 91 (2019). 

\bibitem{kaku1993}
M.\ Kaku,
{\it Quantum Field Theory: A Modern Introduction},
(Oxford University Press, New York, Oxford, 1993).

\bibitem{zeldovich1968}
Ya.\ B.\ Zel'dovich,
The Cosmological Constant and the Theory of Elementary Particles,
Sov.\ Phys.\ Uspekhi, 11, 381 (1968).

\bibitem{weinberg1989}
S.\ Weinberg,
The cosmological constant problem,
Rev.\ Mod.\ Phys., 61, 1 (1989).

\bibitem{horvathy2008}
P.\ A.\ Horv\'{a}thy, M.\ S.\ Plyushchay, and M.\ Valenzuela,
Bosonized supersymmetry from the Majorana-Dirac-Staunton theory 
and massive higher-spin fields,
Phys.\ Rev.\ D 77, 025017 (2008).

\bibitem{okninski2014}
A.\ Okni\'{n}ski,
On the Mechanism of Fermion-Boson Transformation,
Int.\ J.\ Theor.\ Phys., 53, 2662 (2014).

\bibitem{nayak1996}
C.\ Nayak and F.\ Wilczek, 
2n-quasihole states realize $2^{n-1}$-dimensional spinor braiding 
statistics in paired quantum Hall states,
Nucl.\ Phys.\ B479, 529 (1996).

\bibitem{wilczek1998}
F.\ Wilczek, 
Projective Statistics and Spinors in Hilbert Space,
arXiv:hep-th/9806228

\bibitem{karpilovsky1985}
G.\ Karpilovsky,
{\it Projective Representations of Finite Groups}
(Marcel Dekker, Inc., New York, 1985). 

\bibitem{hoffman1992}
P.\ N.\ Hoffman and J.\ F.\ Humphreys,
{\it Projective Representations of the Symmetric Groups}
(Clarendon Press, Oxford, 1992). 

\bibitem{baugh2001}
J.\ Baugh, D.\ Finkelstein, A.\ Galiautdinov, H.\ Saller,
Clifford algebra as quantum language,
J.\ Math.\ Phys., 42, 1489 (2001).

\bibitem{finkelstein2001}
D.\ R.\ Finkelstein and A.\ A.\ Galiautdinov,
Cliffordons,
J.\ Math.\ Phys., 42, 3299 (2001).

\bibitem{galiautdinov2002a}
A.\ A.\ Galiautdinov and D.\ R.\ Finkelstein,
Chronon corrections to the Dirac equation,
J.\ Math.\ Phys., 43, 4741 (2002).

\bibitem{galiautdinov2002b}
A.\ A.\ Galiautdinov,
Quantum Theory of Elementary Processes,
Int.\ J.\ Theor.\ Phys., 41, 1423 (2002).

\bibitem{baugh2003}
 J.\ Baugh, D.\ R.\ Finkelstein, A.\ Galiautdinov, and M. Shiri-Garakani,
Transquantum Dynamics,
Found.\ Phys., 33, 1267 (2003).

\bibitem{marks2022}
D.\ W.\ Marks,
Binary Encoded Recursive Generation
of Quantum Space-Times,
Adv. Appl. Cliff.\ Alg., 32, 51 (2022).

\bibitem{galiautdinov2002c}
A.\ A.\ Galiautdinov,
Clifford statistics and the temperature limit in the theory of fractional quantum Hall effect,
arXiv:hep-th/0201052

\bibitem{read2003}
N.\ Read,
Non-Abelian braid statistics versus projective
permutation statistics,
J.\ Math.\ Phys., 44, 558 (2003).


\bibitem{duval2001}
C.\ Duval and P.\ A.\ Horv\'{a}thy,
Exotic Galilean symmetry in the non-commutative plane and the Hall effect,
J.\ Phys.\ A: Math.\ Gen., 34, 10097 (2001).

\bibitem{horvathy2002}
P.\ A.\ Horv\'{a}thy and M.\ S.\ Plyushchay, 
Non-relativistic anyons, exotic Galilean symmetry and noncommutative plane,
JHEP, 06, 033 (2002).

\bibitem{horvathy2004}
P.\ A.\ Horv\'{a}thy and M.\ S.\ Plyushchay, 
Anyon wave equations and the noncommutative plane,
Phys.\ Lett.\ B 595, 547 (2004).

\bibitem{levy-leblond1967}
J.-M.\ L\'{e}vy-Leblond, 
Nonrelativistic Particles and Wave Equations,
Commun.\ Math.\ Phys., 6, 286 (1967).

\bibitem{davies1977}
P.\ C.\ W.\ Davies, 
{\it The Physics of Time Asymmetry} 
(University of California Press, Berkeley and Los Angeles, 1977).

\bibitem{penrose1979}
R.\ Penrose,
Singularities and time-asymmetry, in {\it General relativity: an Einstein
centenary survey}, edited by S.\ W.\ Hawking and W.\ Israel 
 (Cambridge University Press, 1979).

\bibitem{hawking1985}
S.\ W.\ Hawking, 
The Arrow of Time in Cosmology, 
Phys.\ Rev.\ D 32, 2489 (1985). 


\bibitem{gell-mann1994}
M.\ Gell-Mann, J.\ B.\ Hartle,
Time Symmetry and Asymmetry in Quantum Mechanics and 
Quantum Cosmology, in {\it Physical Origins of Time Asymmetry}, 
ed.\ by J.\ Halliwell, J.\ Perez-Mercader, and W.\ Zurek 
(Cambridge University Press, Cambridge, 1994).

\bibitem{zeh2011}
H.\ D.\ Zeh, 
Open questions regarding the arrow of time, in 
{\it The Arrows of Time} (Springer, Berlin, Heidelberg, 2011). 

\bibitem{ellis2013}
G.\ F.\ R.\ Ellis, 
The arrow of time and the nature of spacetime,
Stud.\ Hist.\ Phil.\ Sci., B44, 242 (2013).

\bibitem{vilenkin2013}
A.\ Vilenkin,
Arrows of time and the beginning of the universe,
Phys.\ Rev.\ D 88, 043516 (2013).

\bibitem{cortes2018}
M.\ Cortes and L.\ Smolin,
Reversing the irreversible: From limit cycles to emergent 
time symmetry,
Phys.\ Rev.\ D 97, 026004 (2018).
 
\bibitem{mandelstam1962}
S.\ Mandelstam,
Quantum Electrodynamics Without Potentials,
Ann.\ Phys., 19, 1 (1962). 

\end{thebibliography}
\end{document}